\documentclass[tightenlines,11pt,eqsecnum,amsmath,amssymb,aps,prd,superscriptaddress,nofootinbib]{revtex4}
\usepackage{amstext,amsfonts}
\usepackage{epsfig}
\usepackage{color}
\usepackage{hyperref}
\usepackage{graphics}
\usepackage{footnote}

\begin{document}
\title{A dynamical system representation of generalized Rastall gravity}
\author{Hamid Shabani}\email{h.shabani@phys.usb.ac.ir}\affiliation{Physics Department, Faculty of Sciences, University of Sistan and Baluchestan, Zahedan, Iran}\affiliation{School of Astronomy, Institute for Research in Fundamental Sciences (IPM)
P. O. Box 19395-5531, Tehran, Iran}
\author{Hooman Moradpour}\email{hn.moradpour@maragheh.ac.ir}\affiliation{Research Institute for Astronomy and Astrophysics of Maragha (RIAAM), University of Maragheh, P.O. Box 55136-553, Maragheh, Iran}
\author{Amir Hadi Ziaie}\email{ah.ziaie@maragheh.ac.ir}\affiliation{Research Institute for Astronomy and Astrophysics of Maragha (RIAAM), University of Maragheh, P.O. Box 55136-553, Maragheh, Iran}
%\date{August 18, 2013}
%\preprint{hep-th/yymmnnn}
%
\begin{abstract}
In this work we study the phase-space analysis of generalized Rastall gravity (GRG) which has recently been introduced as a modification to the original version of Rastall gravity (RG). In GRG, the coupling parameter assumes a dynamical feature and may play the role of dark energy (DE) which is responsible for the present accelerating expansion of the Universe. Our investigation shows that such a modification of General Relativity (GR) admits a stable critical point corresponding to the late time accelerated expansion of the Universe. Assuming both dark matter (DM) and an ultra-relativistic perfect fluid (radiation) as the cosmic ingredients, we find that the underlying model presents a viable sequence of cosmic evolution, beginning from radiation dominated era passing then through DM domination and finally reaches the late time DE dominated era. Moreover, taking into account the contribution due to the spatial curvature within the total energy density leads to a growing mode for it in such a way that its present values are consistent with those reported in recent observations. We also present a numerical simulation of the model which particularizes our dynamical system representation. These studies demonstrate the ability of GRG to describe the present accelerated expansion of the Universe. 
\end{abstract}

%\pacs{04.50.Kd; 95.36.+x; 98.80.-k; 98.80.Jk}
% \keywords{Cosmology; $f(R,T)$ Gravity; Dark Energy; Dynamical Systems Approach; Modified Theories of Gravity.}
\maketitle
\section{Introduction}\label{Int}
It is now known that {GR} is the most successful and accurate gravitational theory at classical level. Its prominent description of the gravitational interaction as a purely manifestation of spacetime geometry along with numerous experimental evidences has exalted it as the backbone of modern theory of gravitational interactions, relativistic cosmology and astrophysics~\cite{Willbook}. Since its advent, the theory has passed a series of rigorous tests ranging from the solar system~\cite{GRtest} to the larger scales such as gravitational lensing by distant galaxies and super-clusters~\cite{bartelmann2010}. Even up until today, its basic foundations and further implications are continually being reviewed and examined, as in the case of the recent discovery of gravitational waves from a binary black hole system~\cite{BBlackhole} and the first direct experimental verification of the existence of black holes in the Universe~\cite{EHTP}. 
\par
Despite its successes, GR suffers from some shortcomings, particularly in the cosmic large scales. This is why alternative theories as extensions of GR have always attracted much attention due to the deep related principal concepts and open issues still unanswered by {GR} such as the problem of invisible components of gravitating matter, i.e., DE and DM~\cite{ExtGravCappo}. Also, two famous problems are the inability of GR to justify the initial and final accelerated expansion states during the evolution of the Universe which have formed the basis for a rich literature, up to now. The former was introduced to solve some fundamental problems, e.g., the flatness and horizon problems~\cite{weinberg2008} and the latter is confirmed by recent observations e.g., type Ia supernovae (SNIa) observations~\cite{supno1,supno2,supno3,supno4,supno5}, baryon acoustic oscillations (BAO)~\cite{BAO1,BAO2,BAO3}, weak lensing~\cite{wlen1}, large-scale structure (LSS)~\cite{Lss1,Lss2}, and the cosmic microwave background radiation (CMBR)~\cite{CMBR1,CMBR2,CMBR3}. In order to tackle with these issues different approaches are examined most of which are based on modifications or generalizations of the parent theory, i.e., GR. Some of these ideas include extra dimensions such as Kaluza-Klein theories~\cite{kaluza} and brane world scenarios~\cite{brane}. In some of them the gravitational coupling constant is taken as a dynamical scalar field e.g., Brans-Dicke theory~\cite{brans}. Higher order gravities have also been introduced as suitable generalizations of GR among which $f(R)$ gravity is the most famous ones~\cite{farhoudi2006,nojir2007,sotiriou2007,defelice2010,sotiriou2010,nojir2011,clifton2012}. Recently, $f(R,T)$ gravity, introduced as a generalization of $f(R)$ gravity, has attracted much attention in scientific community. In this theory an arbitrary function of Ricci curvature scalar along with the trace of Energy momentum tensor (EMT) is used to modify the Einstein-Hilbert action~\cite{shab2013,harko2014,zare2016,shab20171,shab20172,shab20173,
shab20174,shab20181,shab20182,singh2018,nagpal2019,sharif2019,moraes2019,ordines2019,
bhatta2019,elizalde2019,baffou2019,bhar2021}. Another approach is to introduce new material ingredients to GR field equation, namely, an exotic type of matter called DM~\cite{bertonea2005, silk2006,feng2010,frenk2012,bergstrom2012} which is necessary in formation and evolution of large scale structures as well as explaining the galaxy rotation curves and an unknown energy component called DE~\cite{peebles2003,polarski2006,copeland2006,durrer2008,bamba2012,bharali2021,sardar2021} which is responsible for the present accelerated expansion of the Universe.
\par
Most of alternative gravity theories have been created to answer questions about the consistency of GR with what we see in the Universe in the present time. These theories satisfy the conservation of EMT which is expressed by $T^{\mu\nu}_{~~;\mu}=0$. In 1972 Rastall confronts the scientific community with a very interesting question~\cite{rastall1972}. Despite of accepted concept based on some physical implications, why should the conservation of EMT hold in a generally curved spacetime? The main reason of Rastall for posing such a challenge was that this convention has been verified only in the special relativity regime or in the limit of weak
gravitational fields and there have not yet been strong evidences for conservation of EMT to hold generally. The principle of equivalence can be accounted for the only conceptual (and so non-experimental) supporter of this presumption~\cite{rastall1972}. Historically, then he considered $T^{\mu\nu}_{ ;\mu}=a^{;\nu}$ as a modification to EMT conservation in curved spacetimes provided that the vector field $a_\alpha$ vanishes in a flat spacetime. His later version of this presupposition simply was to constructing a vector field using the Ricci scalar as, $a_\mu=\lambda' R_{;\mu}$ for an arbitrary constant $\lambda'$ (which is called the Rastall parameter). This choice leads to the following field equation
\begin{eqnarray}\label{grI1}
{\sf G}_{\mu\nu}+\lambda'\kappa'{\sf R}g_{\mu\nu}=\kappa'{\sf T}_{\mu\nu}\nonumber,
\end{eqnarray}
where, $\kappa'$ is the gravitational coupling in this theory. This scenario is known as RG in the literature and has been widely investigated till now. Among many studies in this area some of them can be highlighted; the cosmological implications of RG have been considered in~\cite{batista2010,capone20101,capone20102,fabris2011,batista2012,batista2013,silva2013}; in~\cite{santos2015} G\"{o}del-type solution has been studied, inclusion of the Brans-Dicke scalar field is studied in~\cite{thiago2014}. The authors of~\cite{majernik2006} have considered compatibility of RG with Mach's principle, in~\cite{oliveira2015,oliveira2016,lin20191} static spherically symmetric solutions have been presented and in~\cite{ziaie2021} effects of Rastall parameter on DE perturbations has been studied. In search of constructing a Lagrangian formulation of RG, the authors of~\cite{moraes2019} have found that some modified theories of gravity can result in RG in some particular cases. Other investigations have been performed in~\cite{yuan2016,darabi2018,hansraj2019,halder2019,khyllep2019,yu2019,dhruba2021}.\\

Recently, an extension of RG has been introduced in which the Rastall parameter is allowed to be a variable~\cite{moradpour20171} (in this case $a_\mu=(\lambda R)_{;\mu}$ where $\lambda$ is the running Rastall parameter). It has been shown that this scenario (which we call it ``the generalized Rastall gravity" (GRG)) is enable to resolve an important problem of RG and also GR, i.e., the accelerated expansion phase of the Universe. More precisely, in~\cite{moradpour20171} the authors showed that in a flat FLRW background, GRG is capable of explaining an inflationary phase even without a matter component. Also, the existence of pre- and post-inflationary solutions in GRG has been reported in~\cite{das2018}. In a different work the authors have extended RG exploiting an arbitrary function of $R$, i.e., $f(R)$ instead of $R$ itself~\cite{lin20192}. A possible connection between GRG and $f(R,T)$ gravity has been also studied~\cite{shabani2020}. Recently, in an interesting study the authors of~\cite{moradpour2021} have demonstrated that GRG presents a framework to allow for a variable Newtonian gravitational coupling, $G$, and thus a compatibility with Dirac hypothesis~\footnote{Dirac hypothesis simply state the idea that $G$ may have been a variable instead of a constant in the whole cosmic eras.}; particularly the corresponding solution to the current accelerated expansion of the Universe can be achieved.\\
Since the GRG scenario is still a young model, it has been not extensively investigated. Particularly, because of arbitrariness of the Rastall parameter, it deserves a deep study in various scopes. In the present work we utilize the dynamical system approach (DS) in order to provide a survey for evolution of the Universe, assuming a specific form for the Rastall parameter $\lambda$. Obviously, other similar solutions may still exist corresponding to another forms of the Rastall parameter. In Sect.~\ref{sec2} we write the GRG field equations for a determined form of the Rastall parameter. For a more complete study the spatial curvature and the ultra-relativistic matter contributions are also included. In Sect.~\ref{sec2} we proceed with rewriting all necessary equations in terms of some dimensionless variables which are defined in the same section. Sect.~\ref{sec3} is devoted to study the solutions of the autonomous system, introduced in the previous section. To be more precise, numerical investigation on the field equations has be presented in Sect.~\ref{sec4} for the same Rastall parameter. Finally, in Sect.~\ref{sec5} we summarize the results. In this work we set the units so that $c=\hbar=8\pi G=1$.
%%%%%%%%%%%%%%%%%%%%%%%%%%%%%%%%%%%%%%%%%%
\section{The field equations of GRG and the corresponding dimensionless variables}\label{sec2}
In this section we briefly present the GRG field equations and all necessary equations. Also, we define the dimensionless variables which we need throughout the paper. The GRG presumes the following constraint on the EMT
\begin{eqnarray}\label{gr0}
T^{\mu\nu}_{\ \ \ ;\mu}=\left(\lambda R\right)^{;\nu},
\end{eqnarray}
where $T_{\mu\nu}$ and $R$ denote the EMT and the Ricci curvature scalar, respectively. Also, the Rastall parameter, $\lambda$, which is a measure of mutual interaction between matter and geometry, is a varying parameter that depends on spacetime coordinates. In the framework of RG, the Rastall parameter is a constant and thus play a minor role in the evolution of the Universe. However, as we shall see, in GRG the presence of a varying coupling parameter could provide reasonable and interesting scenarios for evolution of the Universe.

Employing Bianchi identities for Einstein tensor, i.e., $G^{\mu\nu}_{\ \ \ ;\mu}=0$ along with using Eq.~(\ref{gr0}), leaves us with the field equation of GRG, given as
\begin{eqnarray}\label{gr1}
G_{\mu\nu}+\kappa\lambda g_{\mu\nu}R=\kappa T_{\mu \nu},
\end{eqnarray}
where $\kappa$ is an integration constant which is called the Rastall gravitational coupling constant. In \cite{moradpour2021}, it has been shown that a particular form of $\lambda$ parameter allows for Dirac proposal on large dimensionless numbers which is well treated in the framework of GRG. Motivated by this result, in the present work, we consider the same form for the functionality of $\lambda$ parameter. Note that here the gravitational coupling constant does not vary with time and thus the present study is completely different from the one which has been presented in~\cite{moradpour2021}. We therefore take the following function for the Rastall parameter
\begin{eqnarray}\label{gr2}
\kappa\lambda=\zeta\frac{f(H)}{R},~~~~~~~f(H)=H^{n},
\end{eqnarray}
where $\zeta$ is an arbitrary constant and $f(H)$ indicates an arbitrary function of the Hubble parameter. For matter content, we assume the EMT of a perfect fluid, given by
\begin{align}\label{gr3}
T_{\mu\nu}=(\rho+p)u_\mu u_\nu + p g_{\mu\nu},
\end{align}
where $\rho$, $p$ and $u_\alpha$ are the matter density, its pressure and the velocity four-vector of the fluid, respectively. The line element for a homogeneous and isotropic Universe is parameterized by the FLRW metric
\begin{align}\label{gr4}
ds^{2}=-dt^{2}+a^{2}(t) \left [\frac{dr^{2}}{1-kr^2}+r^{2}d\Omega^2\right ],
\end{align}
where $a(t)$, $k$ and $d\Omega^2$ denote the scale factor, the spatial curvature constant and the line element on a unit two-sphere, respectively. For the above metric, the non-vanishing components of field equation (\ref{gr1}) are obtained as
\begin{align}
&3H^2 - \zeta f(H)+3\frac{k}{a^{2}}=\kappa \left(\rho^{\rm(dm)}+\rho^{\rm(rad)}\right)\label{gr5},\\
&2\dot{H}+3H^2 - \zeta f(H) + \frac{k}{a^{2}}=-\kappa\frac{\rho^{\rm (rad)}}{3}\label{gr6},
\end{align}
where use has been made of assumption (\ref{gr2}) and $\rho^{\rm (dm)}$ and $\rho^{\rm (rad)}$ denote the DM and ultra-relativistic matter densities, respectively. We continue with this assumption that these two fluids evolve independently and since the Ricci scalar vanishes for ultra-relativistic matter, the matter-curvature coupling only affects the DE evolution.
\par
To rewrite Eqs.~(\ref{gr5}) and (\ref{gr6}) as a closed dynamical system we must define adequate dimensionless variables
which can be determined using Eq.~(\ref{gr5}) as a constraint equation. We then get
\begin{align}\label{gr7}
\kappa\frac{\rho^{\rm(dm)}}{3H^2}+\kappa\frac{\rho^{\rm(rad)}}{3H^2}+\zeta \frac{f(H)}{3H^2}-\frac{k}{a^{2}H^{2}}=1,
\end{align}
from which we can define the following dimensionless variables and constants
 \begin{align}\label{gr8}
&\Omega^{\rm(dm)}=\frac{\kappa_{\rm G}\rho^{\rm(dm)}}{3H^2},~~~~~\Omega^{\rm(rad)}=\frac{\kappa_{\rm G}\rho^{\rm(rad)}}{3H^2},~~~~~\Omega^{(k)}=-\frac{k}{a^{2}H^{2}},\\
&\Omega^{\rm(de)}=\frac{\zeta f(H)}{3H^2},~~~~~\alpha=\frac{\kappa}{\kappa_{\rm G}},~~~~~\mathcal{M}=\frac{f'(H)H}{f(H)}\nonumber,
\end{align}
where the prime denotes differentiation with respect to the Hubble parameter and $\kappa_{\rm G}=8\pi G$=1. In the case of ansatz (\ref{gr2}), we have $\mathcal{M}=n$. Therefore, in terms of the above definitions, Eq.~(\ref{gr5}) takes the following form
\begin{align}\label{gr9}
\alpha\Omega^{\rm (dm)}+\alpha\Omega^{\rm (rad)}+\Omega^{\rm (de)}+\Omega^{(k)}=1.
\end{align}
\\
Thus, the closed dynamical system corresponding to Eqs.~(\ref{gr5}) and (\ref{gr6}) reads
 \begin{align}
&\frac{d\Omega^{\rm (rad)}}{d N}=\Omega^{\rm (rad)}\left(\alpha\Omega^{\rm (rad)}-\Omega^{(k)}-3\Omega^{\rm (de)}-1\right ),\label{gr10}\\
&\frac{d\Omega^{(k)}}{d N}=\Omega^{(k)}\left(\alpha\Omega^{\rm (rad)}-\Omega^{(k)}-3\Omega^{\rm (de)}+1\right ),\label{gr11}\\
&\frac{d\Omega^{\rm (de)}}{dN}=\left(\frac{n}{2}-1\right)\Omega^{\rm (de)}\left(-\alpha\Omega^{\rm (rad)}+\Omega^{(k)}+3\Omega^{\rm(de)}-3\right ),\label{gr12}
\end{align}
where $N$ is defined so as to have $Hdt=dN$. Also, Eq.~(\ref{gr6}) is reformulated as
\begin{align}\label{gr13}
\frac{2\dot{H}}{3H^{2}}=-\frac{\alpha}{3}\Omega^{\rm(rad)}+\frac{\Omega^{(k)}}{3}+\Omega^{\rm(de)}-1,
\end{align}
where we have used Eq.~(\ref{gr9}). Using Eq.~(\ref{gr13}) one can obtain the equation of state parameter (ESP) $w$ and the deceleration parameter (DP) $q$ as follows
\begin{align}
&w=\frac{\alpha }{3}\Omega^{\rm (rad)}-\Omega^{\rm (de)}-\frac{\Omega^{(k)}}{3},\label{gr14}\\
&q=\frac{\alpha }{2}\Omega^{\rm (rad)}-\frac{3}{2} \Omega^{\rm (de)}-\frac{\Omega^{(k)}}{2}+\frac{1}{2}.\label{gr15}
\end{align}
In the next section we investigate the autonomous system (\ref{gr10})-(\ref{gr12}) and seek for an acceptable cosmological scenario.
 %%%%%%%%%%%%%%%%%%%%%%%%%%%%%%%%%%%%%%%%%%%%%%%%%%%
\section{The dynamics of dimensionless variables}\label{sec3}
In the present section we inspect the behavior of dynamical variables $\Omega^{(k)}$, $\Omega^{\rm(rad)}$, $\Omega^{\rm(de)}$ as well as ESP and DP, utilizing Eqs.~(\ref{gr10})-(\ref{gr12}) along with (\ref{gr14})-(\ref{gr15}). The fixed point solutions of system (\ref{gr10})-(\ref{gr12}) for arbitrary value of parameter $n$ are listed in Table~(\ref{tab}). For each fixed point, the corresponding ESP and DP has been calculated. Note that, to avoid any ambiguity we continue with constant values of $n$.
\begin{center}
\begin{table}[h!]
\centering
\caption{The fixed point solutions of GRG with {Rastall parameter} (\ref{gr2}).}
\begin{tabular}{l @{\hskip 0.1in} l@{\hskip 0.1in} l @{\hskip 0.1in}l @{\hskip 0.1in}r@{\hskip 0.1in}r}\hline\hline

Fixed point     & $(\Omega^{\textrm{(k)}},\Omega^{\textrm{(rad)}},\Omega^{\textrm{(de)}})$           &Eigenvalues  &$\Omega^{\rm(dm)}$      &$w$&$q$\\[0.5 ex]
\hline
$P^{\textrm{(de)}}$&$(0,0,1)$&$\left[-1,-2,\frac{3}{2} \left(-2 + n\right)\right]$&$0$&$-1$&$-1$\\[0.75 ex]
$P^{\textrm{(k)}}$&$(1,0,0)$&$\left(-2, -1, 2-n\right)$&$0$&$-\frac{1}{3}$&$0$\\[0.75 ex]
$P^{\textrm{(rad)}}$&$(0,\frac{1}{\alpha},0)$&$\left[1,2,2\left(2 - n\right)\right]$&$0$&$\frac{1}{3}$&$1$\\[0.75 ex]
$P^{\textrm{(dm)}}$&$(0,0,0)$&$\left[-1,1,\frac{3}{2}\left (2 - n\right)\right]$&$1$&$0$&$\frac{1}{2}$\\[0.75 ex]
\hline\hline
\end{tabular}
\label{tab}
\end{table}
\end{center}

\begin{figure}
\begin{center}
\epsfig{figure=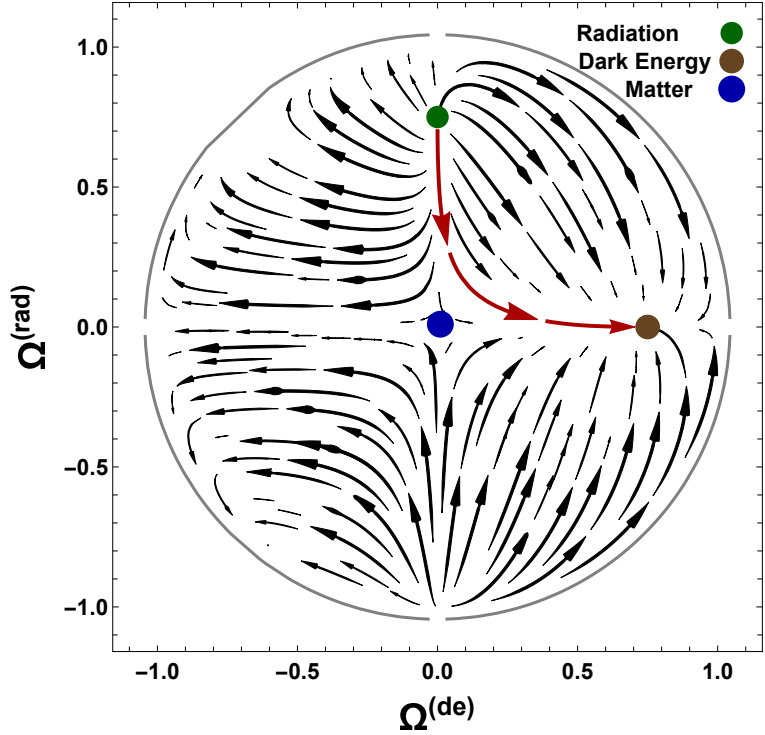,width=10.cm}
\caption{The phase-space portrait of GRG for the fixed point solutions as provided by Table~(\ref{tab}). The Rastall parameter has been taken as (\ref{gr2}) and for model parameters we have set $\Omega^{k}=0$ and $n$=1.}\label{f1}
\end{center}
\end{figure}

From Table~(\ref{tab}), we observe that the four fixed points $P^{\textrm{(de)}}$, $P^{\textrm{(k)}}$, $P^{\textrm{(rad)}}$ and $P^{\textrm{(dm)}}$ correspond only to one dominant density parameter. For $P^{\textrm{(rad)}}$ the density parameter depends on $\alpha$ parameter, therefore, in order to have normalized values, $\alpha$ should be set to unity (hereafter we choose $\alpha$=1). Also, the eigenvalues are linear functions of parameter $n$, which means that it is the type of model (as specified by $n$) that decides the stability of the fixed point; we see easily that only models for which the condition $n<2$ holds include stable DE fixed point, unstable radiation fixed point and two saddle points for the rest ones. Having unstable radiation fixed point is a favorable characteristic which shows that all phase-space trajectories get repelled from this point throughout the  phase-space planes. From cosmological viewpoint, this can be translated into saying that for all models with $n<2$, during its evolution, the Universe experiences a radiation dominated era and then undergoes a transition from this era without ever staying in it. More interestingly, a reverse statement holds for $P^{\textrm{(de)}}$. All models with $n<2$ end up with an accelerated expansion era. The remaining two critical points behave as saddle type solutions which guaranty the enough period of cease for the formation of large scale structures. Figure~(\ref{f1}) shows a phase-space portrait of the fixed point solutions in ($\Omega^{\rm (de)}$,$\Omega^{\rm (rad)}$) plane. We observe that, during its evolution, the Universe passes through the fixed points with physical properties as provided by the pair $(w,q)$ at each point, see the last two columns of Table~(\ref{tab}). For better understanding of the situation, the trajectory of the Universe's evolution has been depicted with red arrows. It is therefore seen that as the Universe expands and cool down, it leaves the radiation era and gradually enters the matter dominated epoch, after which, the Universe begins to be DE dominated and undergoes accelerated expansion. We note that other physical trajectories with overall behavior as of the red arrows are also possible for the evolution of the Universe. However, the difference between these trajectories is that the longer the Universe stays within the matter domination era, the more chance of formation of structures.
\par
\begin{figure}[h]
\begin{center}
\epsfig{figure=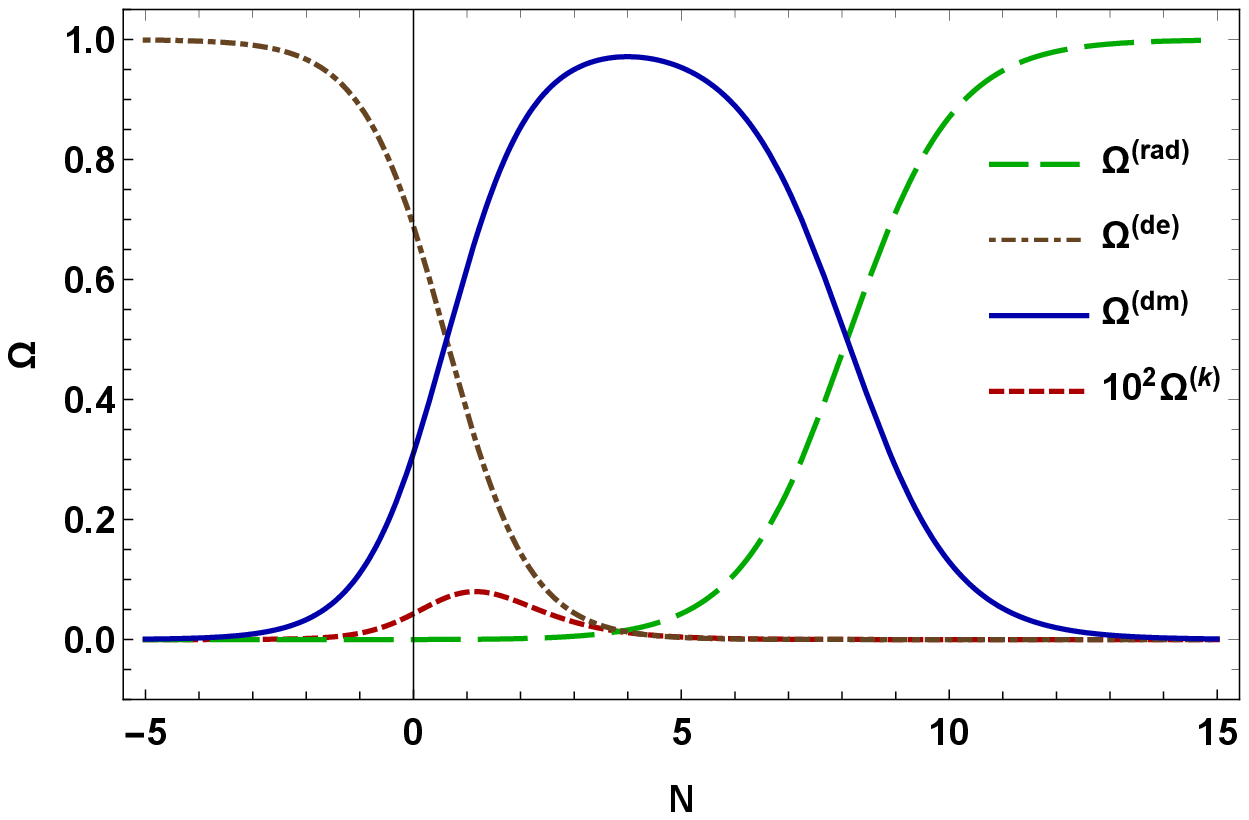,width=8.cm}\hspace{2mm}
\epsfig{figure=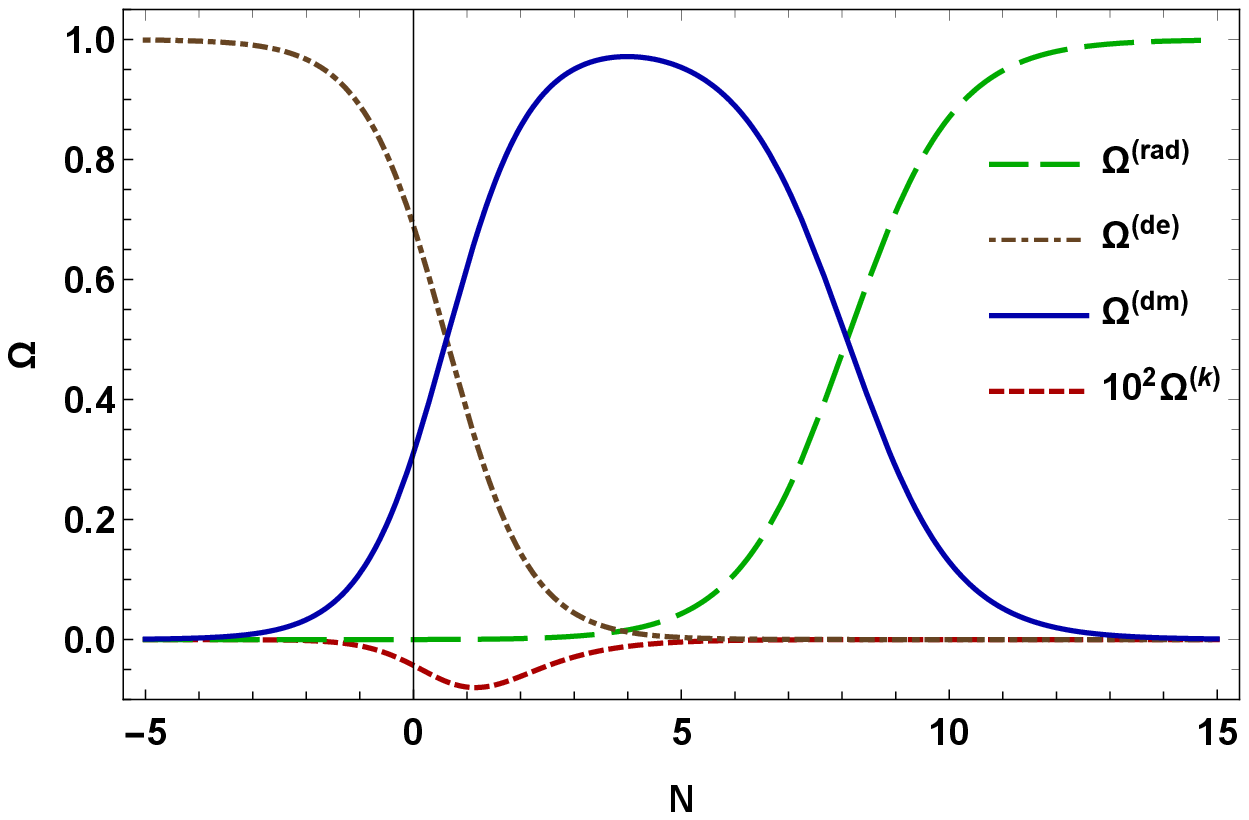,width=8.cm}
\caption{The evolution of different density parameters in GRG for the Rastall parameter (\ref{gr2}) in an open and closed Universe. The initial values $\Omega^{(k)}_{i}=\pm2\times 10^{-12}, \Omega^{\rm(rad)}_{i}=0.999,\Omega^{\rm(de)}_{i}=4.5\times 10^{-10}$ as well as $n=1$ have been set.}\label{f2}
\end{center}
\end{figure}
\begin{figure}[h]
\begin{center}
\epsfig{figure=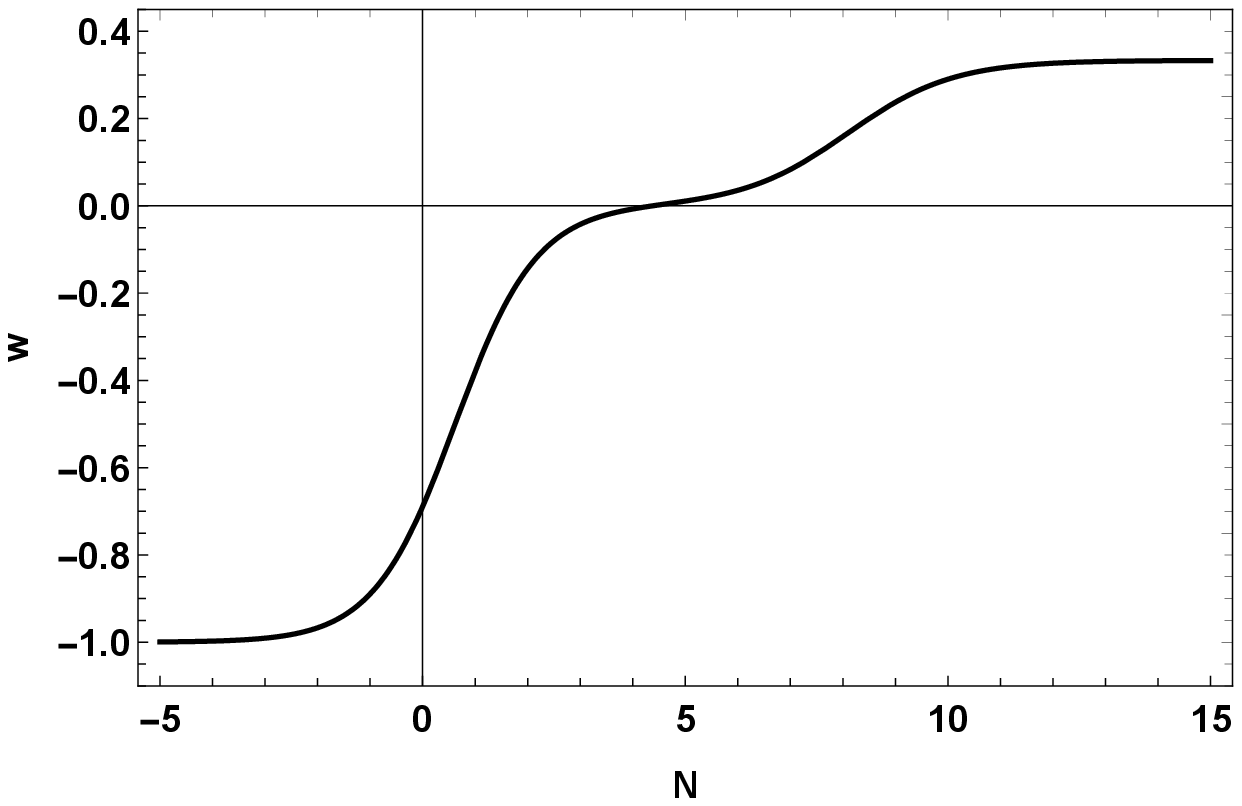,width=8.cm}\hspace{2mm}
\epsfig{figure=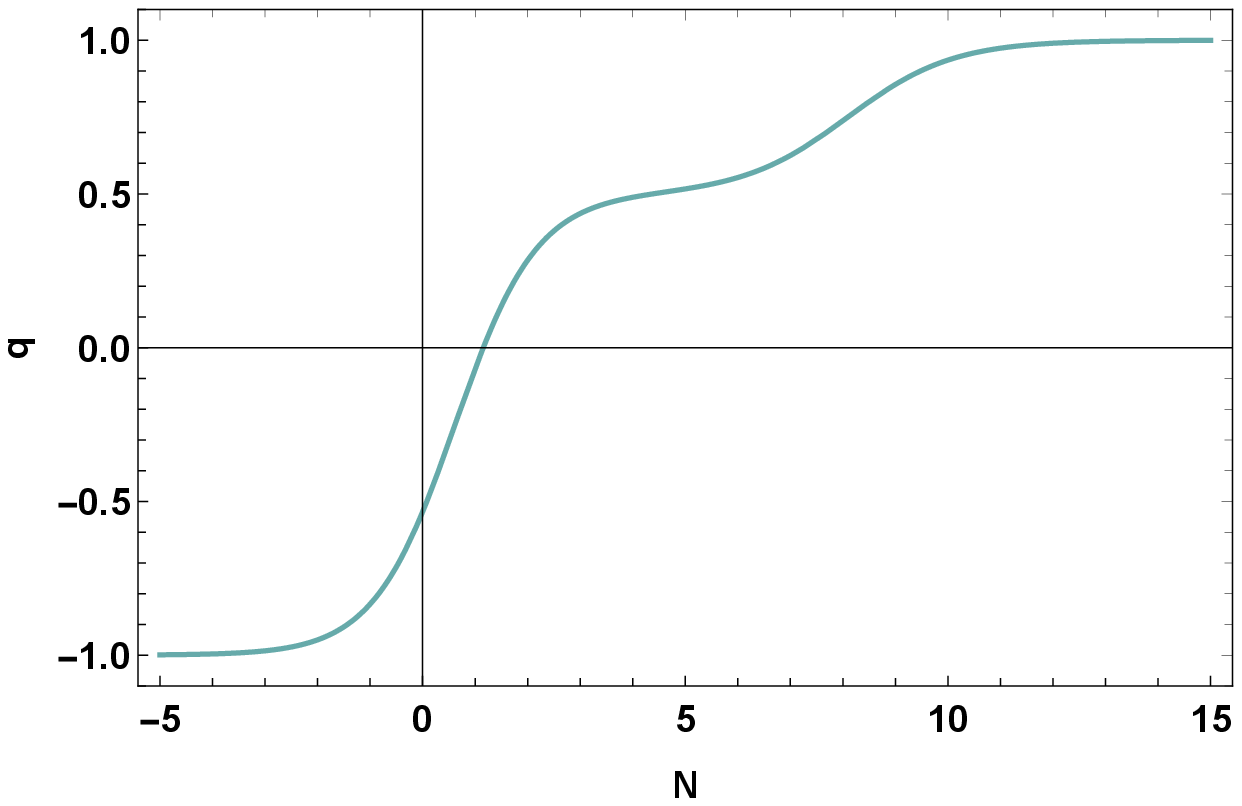,width=8.cm}
\caption{The evolution of ESP and DP for $\Omega^{(k)}_{i}=+2\times 10^{-12}$ and the same initial values as of Fig~(\ref{f2}).}\label{f3}
\end{center}
\end{figure}

In Fig.~(\ref{f2}) we have sketched the dynamical behavior of density parameters in an open and closed Universe. The present values of the radiation and spatial curvature density parameters are found as $\Omega^{\rm (rad)}_{0}\approx 10^{-5}-10^{-4}$, $|\Omega^{(k)}_{0}|\approx 10^{-3}$ which are consistent with the latest observations~\cite{planck2018}. Note that all quantities have been plotted in terms of $N=\ln{(a)}=-\ln{(1+z)}$, thus $N\to0$ corresponds to $z\to0$, which means the present time. 
We have depicted the ESP and DP in Figure~(\ref{f3}) for the same initial values as of Figure~(\ref{f2}). The behavior of these parameters pictures different stages of evolution of the Universe. We observe that a radiation dominated Universe smoothly transits to a situation where DM and baryonic matter are dominant components of the Universe. As time passes, the contribution due to DE begins to grow allowing thus, the Universe to experience an accelerated expanding phase. Both Figures~(\ref{f2}) and~(\ref{f3}) are sketched for $N\approx0.7$ corresponding to $z\approx1$, which is the transition redshift from decelerating to accelerating phases. We also note that the values of density parameters are consistent with the Planck 2018 results~\cite{planck2018}.
%%%%%%%%%%%%%%%%%%%%%%%%%%%Each curve shows the behavior of a density parameter so that all the curves collectively
\section{Numerical Solutions}\label{sec4}
\begin{figure}[h!]
\begin{center}
\epsfig{figure=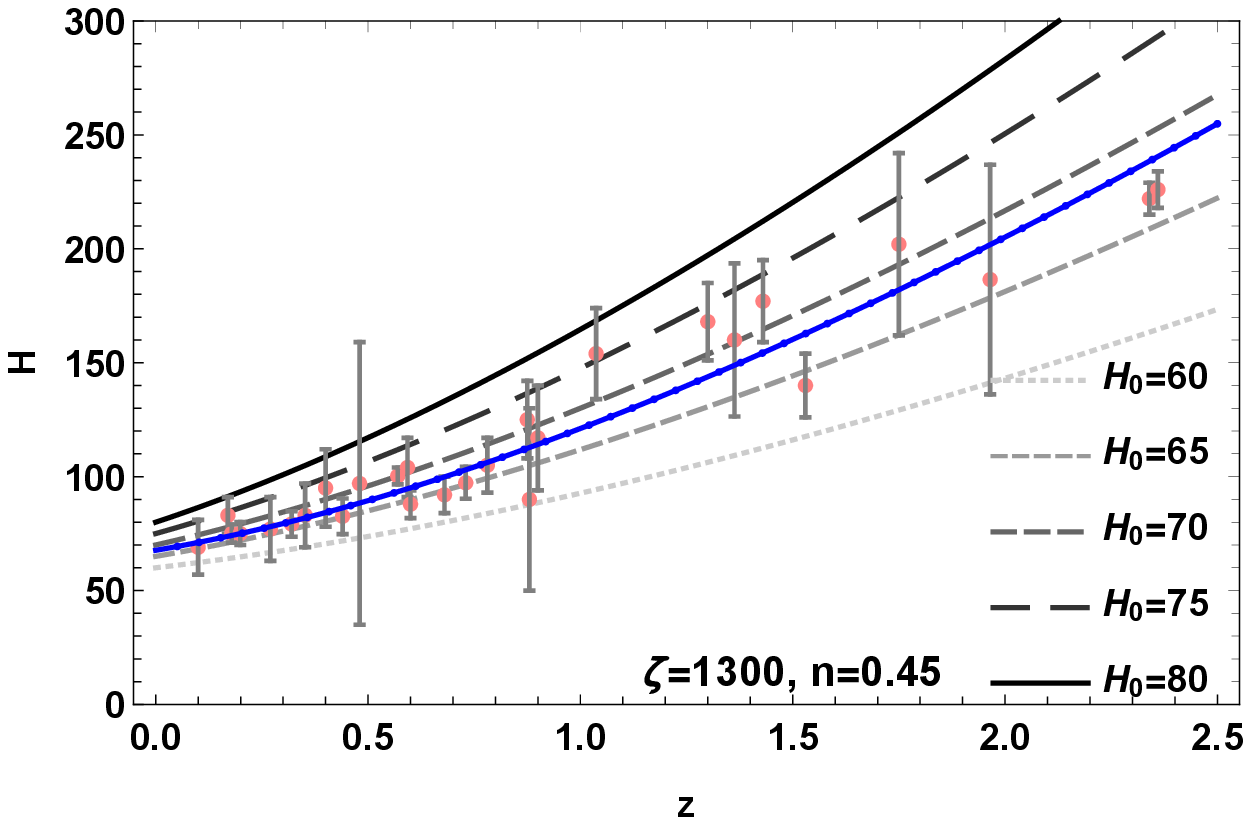,width=8.cm}\hspace{2mm}
\epsfig{figure=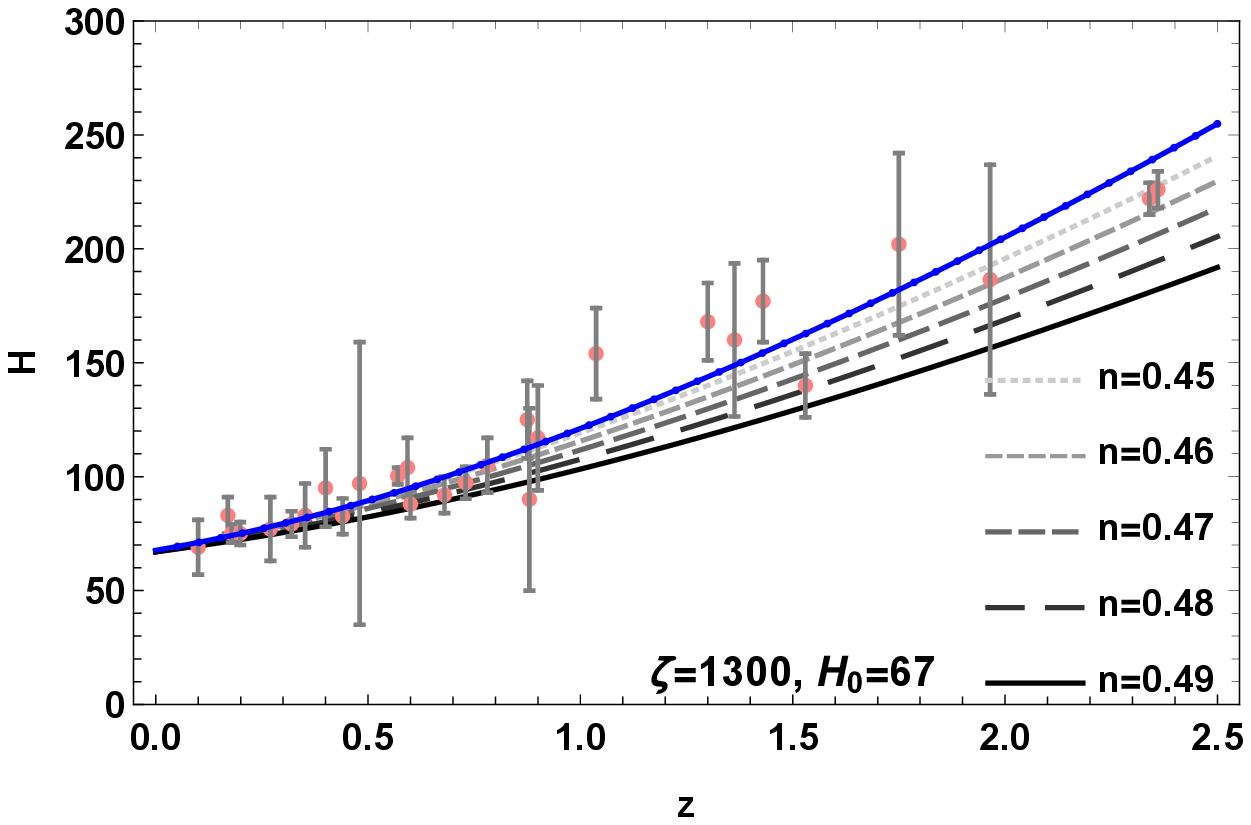,width=8.cm}\vspace{2mm}
\epsfig{figure=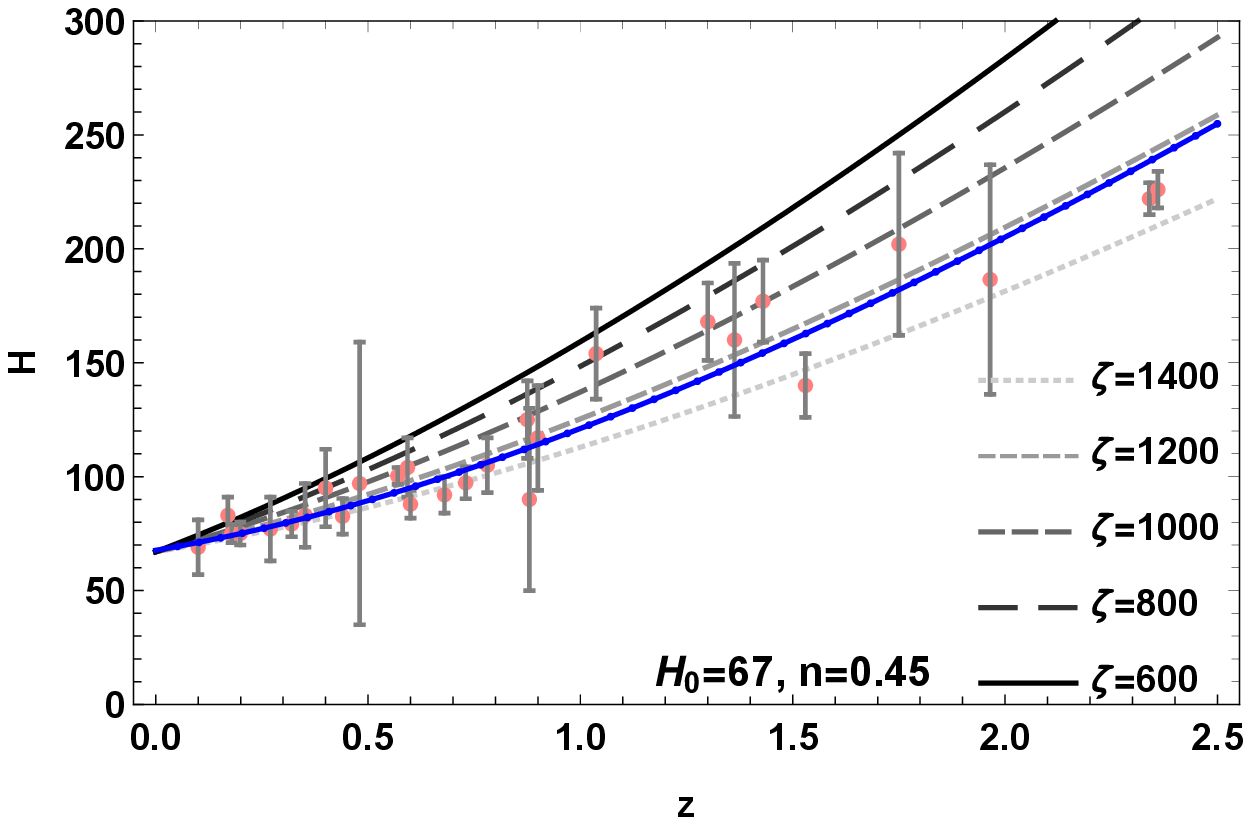,width=8.cm}
\caption{The behavior of Hubble parameter for different astronomical data~\cite{farooq2017}.}\label{f4}
\end{center}
\end{figure}

In the previous section we have discussed general cosmological behavior of GRG model based on a dynamical system point of view assuming the coupling parameter obeys the relation (\ref{gr2}). We observed that in the context of GRG, late time accelerated expansion is achieved. In the present section we investigate numerical simulation of Eqs.~(\ref{gr5}) and (\ref{gr6}) as well as the EMT conservation equation (\ref{gr0}). Considering DM as the only matter component, condition (\ref{gr0}) leads to the following equation
\begin{align}\label{gr16}
\dot{\rho}^{\rm(dm)}+3H\rho^{\rm(dm)}+\zeta f'(H)\dot{H}=0,
\end{align}
where, $f'(H)=df(H)/dH$. Therefore, Eqs.~(\ref{gr6}) and (\ref{gr16}) control the evolution of two variables $H$ and $\rho^{\rm (dm)}$ and Eq.~(\ref{gr5}) acts as a constraint. The exact solution of this system of equations is a complicated inverse  Hypergeometrical function which is not easy to be analyzed. Instead, we discuss numerical solutions. Projecting Eqs.~(\ref{gr5})-(\ref{gr6}) as well as (\ref{gr16}) in redshift space, the behavior of Hubble parameter for various choices of model parameters has been plotted in Fig.~\ref{f4}. Besides, the error bars obtained from observations~\cite{farooq2017} are added to the diagrams for comparison.
In the upper left panel of Figure~\ref{f4} we drawn the Hubble diagrams for $\zeta=1300$ and $n=0.45$ and different values of the Hubble constant ranging from $60~\textrm{km s}^{-1} \textrm{Mpc}^{-1}$ to $80~\textrm{km s}^{-1} \textrm{Mpc}^{-1}$\footnote{These values have been selected to compare to most wide ranges of data. The values $H_{0}=67.6^{+4.3}_{-4.2}~\textrm{km s}^{-1} \textrm{Mpc}^{-1}$ and $H_{0}=75.8^{+5.2}_{-4.9}~\textrm{km s}^{-1} \textrm{Mpc}^{-1}$ have been reported in~\cite{mukherjee2020} and~\cite{jaeger2020} respectively.}. The upper right panel has provided for different values of the model parameter $n$, $\zeta=1300$ and $H_{0}=67$. Also, different Hubble diagrams have been plotted for different values of $\zeta$, $H_{0}=67$ and $n=0.45$. For comparison, the Hubble diagram obtained from GR, the blue dotted solid line, is included in the panels of Fig.~\ref{f4}. We see that the GR curve can be obtained for $\zeta\approx1300$, $H_{0}\approx67$ and $n\approx0.44-0.45$. We used $H_{0}=67$ in this study since the Hubble constant $H_{0} = (67.4 \pm0.5)~\textrm{km s}^{-1} \textrm{Mpc}^{-1}$ has been reported in the Planck data~\cite{planck2018}.\\

\begin{figure}[h!]
\begin{center}
\epsfig{figure=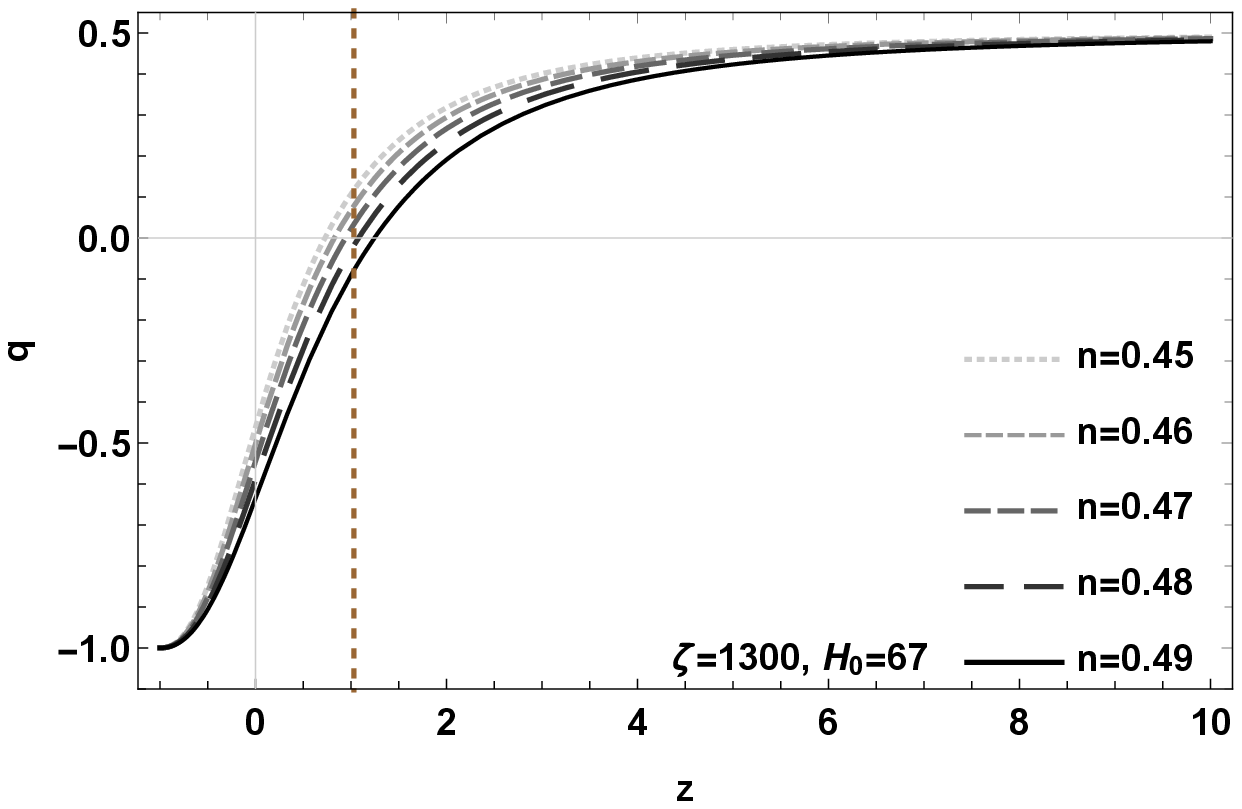,width=8.cm}\hspace{2mm}
\epsfig{figure=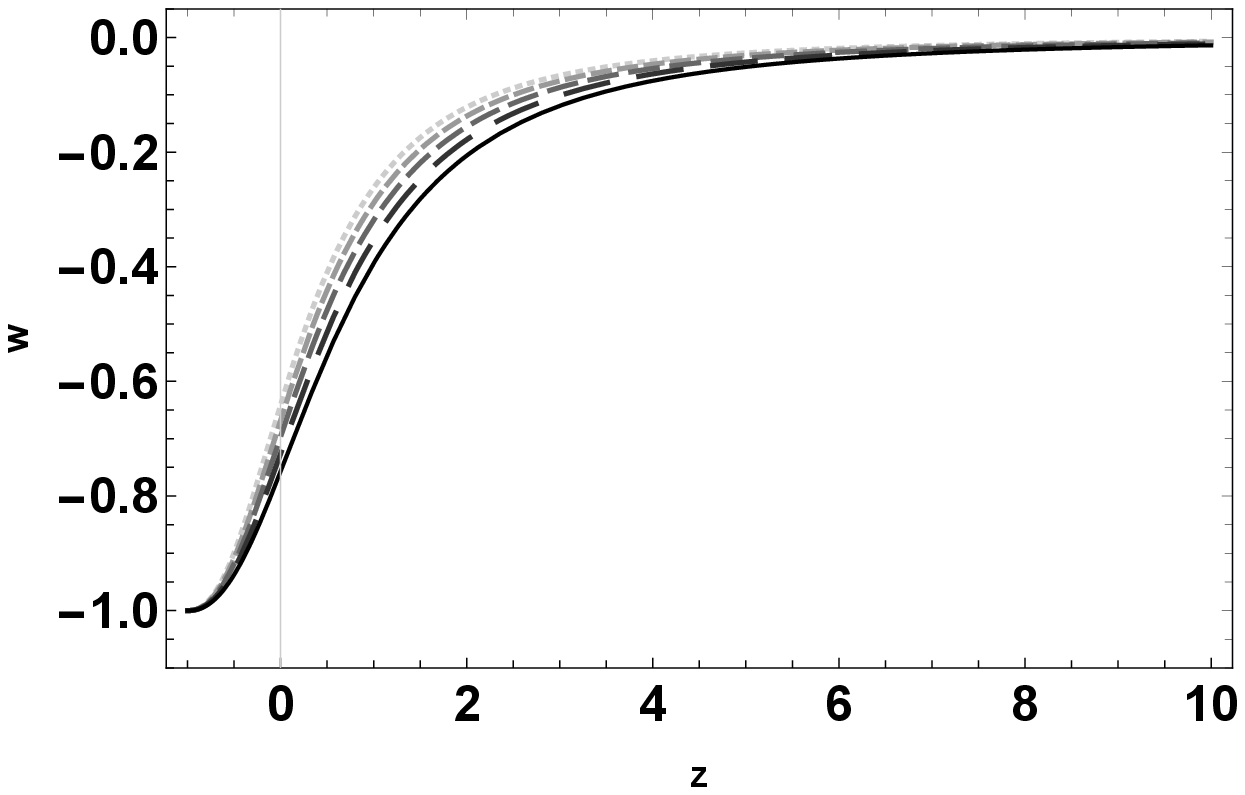,width=8.cm}\vspace{2mm}
\caption{The evolution of DP (left panel) and ESP (right panel) for $\zeta=1300$, $H_{0}=67$ and different values of $n$ parameter. The right panel has been plotted for the same values of $n$ parameter as of the left one.}\label{f5}
\end{center}
\end{figure}

\begin{figure}[h!]
\begin{center}
\epsfig{figure=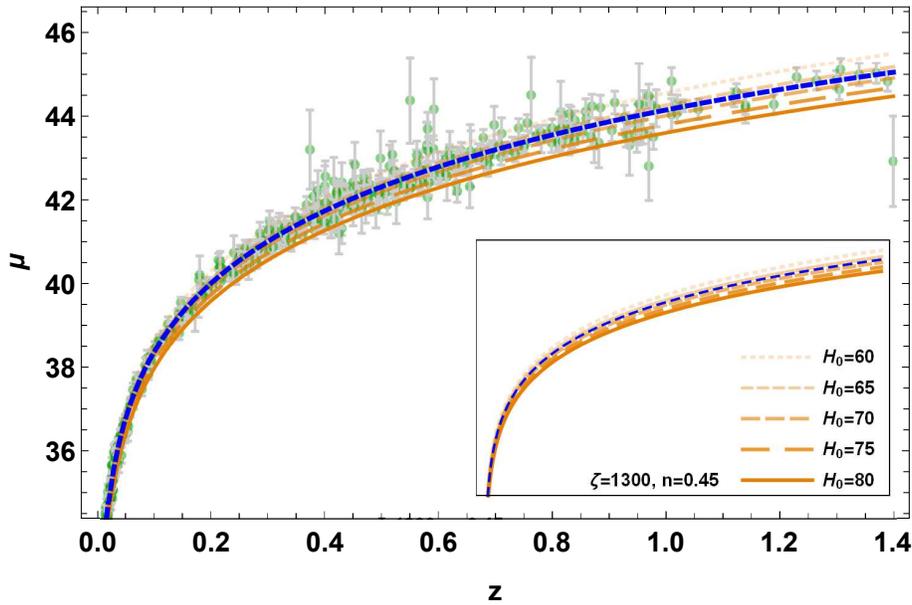,width=12.cm}
\caption{The distance modulus $\mu(z)$ for different values of the Hubble constants. The blue dashed curve shows that of GR. The diagrams have been drawn both with and without the error bars.}\label{f6}
\end{center}
\end{figure}

In Figure~\ref{f5} we have depicted the behavior of DP and ESP for $\zeta=1300$, $H_{0}=67$ and different values of $n$ parameter. Also, the diagram for distance modulus $\mu(z)$ has been provided for $\zeta=1300$, $n=0.45$ and different present values of the Hubble parameter, see Fig.~\ref{f6}. In this figure the dashed blue curve shows the GR distance modulus. The observational data from~\cite{amanullah2010} has been used in Figure~\ref{f6}.
%%%%%%%%%%%%%%%%%%%%%%%%%%%%%%%%%%%%%%%%%%%%%%%%%%%%%%%%%
\section{Concluding Remarks}\label{sec5}
Our aim in the present work was to search for viable cosmological scenarios in the framework of GRG. This theory, is an extension of the original version of Rastall gravity (RG)~\cite{rastall1972} in which, the constant Rastall parameter is replaced by a variable one. The inclusion of such a dynamic parameter, leads to a set of equations of motion (see equations (\ref{gr5}), (\ref{gr6}) and (\ref{gr16})) which are similar to those of GR except an additional term which bears the contribution due to the Rastall parameter. In this work we studied cosmological behavior of GRG based on the choice (\ref{gr2}). Defining some specified dimensionless variables, equations of motion were recast into a set of dimensionless dynamical equations, one of them, depends on parameter $n$. Our investigation showed that, the $n$ parameter affects the stability of the solutions. Our solutions composed of four fixed points which correspond to the ultra-relativistic fluid, the DM, the spatial curvature component and the Rastall parameter term which plays the role of DE. The only attractive fixed point is that of DE dominated era which is stable provided that $n<2$. For $n<2$, the ultra-relativistic fixed point repels all phase-space trajectories and the two other fixed points are of saddle type. Hence, using the dynamical system approach, our study demonstrates that GRG enables one to describe the present accelerated expansion of the Universe, the same result is also reported in~\cite{moradpour20171} utilizing a different method.
\par
We also provided numerical solutions to the field equations and investigated the behavior of DP and ESP in terms of redshift. The behavior of Hubble parameter as well as the distance modulus for different values of the model parameters were discussed. To compare our results, we included the astronomical data error bars within the diagrams. We concluded that for $n\approx0.45$ and $\zeta\approx1300$ and $H_{0}\approx67 \textrm{km s}^{-1} \textrm{Mpc}^{-1}$ our model gets more consistency with the observational data. 

The present work can be accounted for an attempt to inspect the cosmological features of GRG and specifically, the consequences of mutual dynamic interaction between matter and geometry as encoded in the Rastall parameter. Our investigation showed that, though its simplicity, GRG can be regarded as a physically viable theory to model cosmological scenarios.
%

%%%%%%%%%%%%%%%%%%%%%%%%%%%%%%%%%%%%%%%%%%%%%%%%%%%%%%%%

\end{document}